\documentclass
[10pt,aps,prc,twocolumn,showpacs,showkeys,amsmath,floatfix,superscriptaddress]
{revtex4-1}
\usepackage{color,graphicx}
\usepackage{mathptmx}                
\usepackage{dcolumn}                 
\usepackage{bm}                      

\begin{document}

\title{Contribution of the $\rho$ meson and quark substructure to the nuclear spin-orbit potential}

\author{Guy Chanfray}
\affiliation{Univ Lyon, Univ Claude Bernard Lyon 1, CNRS/IN2P3, IP2I Lyon, UMR 5822, F-69622, Villeurbanne, France}

\author{J\'er\^ome Margueron}
\affiliation{Univ Lyon, Univ Claude Bernard Lyon 1, CNRS/IN2P3, IP2I Lyon, UMR 5822, F-69622, Villeurbanne, France}

\begin{abstract}
The microscopic origin of the spin-orbit (SO) potential in terms of sub-baryonic degrees of freedom is explored and discussed  for application to nuclei and hyper-nuclei. 
We thus develop a chiral relativistic approach where the coupling to the scalar- and vector-meson fields 
are controlled by the quark substructure. 
This approach suggests that the isoscalar and isovector density dependence of the SO potential can be used to test the microscopic ingredients  which are implemented in the relativistic framework: the quark substructure of the nucleon in its ground-state and its coupling to the rich meson sector where the $\rho$ meson plays a crucial role. 
This is also in line with the Vector Dominance Model (VDM) phenomenology and the known magnetic properties of the nucleons.
We explore predictions based on Hartree and Hartree-Fock mean field, as well as various scenarios for the $\rho$-nucleon coupling, ranked as weak, medium and strong, which impacts the isoscalar and isovector density dependence of the SO potential. 
We show that a medium to strong $\rho$ coupling is essential to reproduce Skyrme phenomenology in $N=Z$ nuclei as well as its isovector dependence.
Assuming an SU(6) valence quark model
our approach is extended to hyperons and furnishes a microscopic understanding of the quenching of the $N\Lambda$ spin-orbit potential in hyper-nuclei. 
It is also applied to other hyperons, such as $\Sigma$, $\Xi$ and $\Omega$.
\end{abstract}

\date{\today}

\maketitle


Despite its crucial role for the understanding of nuclear magic numbers~\cite{Otto1949,Mayer1950} and consequently of element abundances in our Universe~\cite{Mayer1949}, the microscopic origin of the spin-orbit (SO) interaction is still a matter of discussion.
For practical nonrelativistic nuclear interactions and application to finite nuclei, it is often introduced as a phenomenological term correcting the nuclear interaction~\cite{RingSchuck1984,Bender2003}: it is represented by a short-range interaction describing the coupling of the particle $i$ with spin $\bf{s}_i$ to its orbital angular momentum $\bf{l}_i=\bf{p}_i \times \bf{r}_i$.
As a consequence, the SO potential is characterized by a density gradient term (boost) with isoscalar and isovector contributions, see for instance Refs.~\cite{Reinhard1995,Bender1999} and references therein.

One major success of the relativistic hadrodynamic model initiated by J. Walecka  and coworkers~\cite{SerotWalecka1986,Walecka1997} comes from the natural framework it provides for the SO coupling without the need to introduce an explicit interaction.
The SO interaction originates from the relativistic nature of the hadrodynamic model, often referred to in nuclear physics as the relativistic mean field (RMF) where only the Hartree potential is considered~\cite{Bender2003}, since it is generated by the coupling between the up and down components of the Dirac spinor~\cite{SerotWalecka1986}. 
In particular, a nonrelativistic potential can be derived from the hadrodynamic model showing that the coupling constant of the SO potential
is a function of the meson coupling constants, which are determined from the bulk properties of nuclear matter and/or fits to the nuclear masses. 
The SO splitting appears thus as a prediction of the model since it is not fit \textsl{a priori}.
Its detailed density functional, e.g. isoscalar and isovector density dependence, may however change from one Lagrangian to another.

A functional difference between Skyrme~\cite{Bender2003}  and RMF~\cite{Bender2003} nuclear interaction has been suggested by Reinhard and Flocard~\cite{Reinhard1995}: the Skyrme SO potential combines together isoscalar and isovector density gradient, while the RMF is purely isoscalar.
It should however be noted that the RMF Lagrangian in Ref.~\cite{Reinhard1995} includes only the contribution of the $\sigma$ and $\omega$ mesons to the SO potential.
In Ref.~\cite{Ebran2016} for instance the additional effect of the $\rho$ vector meson and of the Fock term has been discussed, but restricting the $\rho$ to its vector coupling to the nucleon and neglecting the $\rho$-tensor coupling.
In this paper, we explicitly detail the contribution of different mesons to the SO potential and we explore three scenarios for the $\rho$ meson coupling: the weak coupling which neglects the $\rho$-tensor contribution, and the medium and strong coupling which include it with increasing strength. The medium coupling reproduces the pure Vector Dominance Model (VDM) picture~\cite{Bhaduri} while the strong coupling requires an extension of the pure VDM picture and is compatible with the $\pi$ nucleon scattering data~\cite{HP75}.
We show that the $\rho$-tensor coupling and the Fock contribution to the mean field are crucial to reconcile relativistic approaches with Skyrme nuclear phenomenology and more generally to adapt to the experimental data in nuclei and hypernuclei.

In the pure VDM picture the $\rho$-tensor coupling fully contributes to the anomalous magnetic moment of the nucleons. In a microscopically based chiral relativistic approach, this coupling originates from the composite nature of baryons into three quarks. 
Hence, although the SO potential can be seen as a pure relativistic effect, its precise form deeply roots into the nature of the strong (nuclear) interaction, e.g. its chiral realization, the contribution of the scalar and vector meson field dynamics as well as the quark substructure. 
In other words, anchoring the relations between the meson coupling constants into a quark substructured (bag) model, the confrontation of the chiral relativistic predictions for the SO potential to nuclear data can drive to a better understanding of some microscopic aspects of baryons and of their mutual interaction.
There are indeed important questions involving finite nuclei and hyper-nuclei, yet unresolved, which can contribute to the better understanding of the strong interaction and of the quark substructure, such as i) what is the isoscalar and isovector dependence of the SO interaction? and ii) how is the SO interaction modified in hyper-nuclei?
To achieve this program we investigate the role of the quark wave functions in governing the SO coupling of the exchanged mesonic degrees of freedom with the nucleons and the hyperons. We also show the effect of the strange quarks to the SO interaction, providing a deep microscopic understanding of SO splitting in hyper-nuclei.
In the following, our microscopic quark-level derivation of the spin-orbit potential closely follows the one from 
Guichon \textsl{et al.}~\cite{Guichon-so}.

Interestingly the quark substructure of the nucleon impacts also the saturation mechanism of the energy per particle in nuclear matter. Experimentally, the curvature of the energy per particle is directly measured from the energy of the isoscalar giant monopole resonance (ISGMR). It has been suggested that the softening of the equation of state around saturation density is induced by the polarization of the quark internal structure of the nucleons~\cite{Guichon,Chanfray2007,Chanfray2011}. This sub-nucleonic polarization appears in the Lagrangian as a nonlinear meson coupling for the $\sigma$ field or alternatively as a density dependence of the scalar-meson coupling constant. It impacts also the SO potential, but at a subleading order which is neglected in this study. It is however interesting to note that, through the SO interaction and the saturation mechanism, the quark substructure appears to have, at least, two concrete realizations impacting the modeling of the interaction between nucleons.
While it is possible to ignore the microscopic mechanism suggested by the quark substructure in practical nuclear modeling by introducing new terms in the Lagrangian fitted to the properties of finite nuclei, the mechanism we refer to suggests a more global picture which provides a deep understanding of the nature of the nuclear interaction.

This paper is organized as follows: we recall the derivation of the SO potential in atomic and nuclear physics in Sec.~\ref{sec:so:atom}.
We then present a derivation of the nucleonic spin-orbit potential from a chiral Hartree-Fock description in Sec.~\ref{sec:model}.
Our approach is by many aspect based on the one presented in Ref.~\cite{Guichon-so}, and also opens the possibility to perform a fully consistent relativistic Hartree-Fock calculation. The connection to nuclear physics is emphasized in our study.
In Sec.~\ref{sec:nuc} we show the impact of the nucleon substructure in nuclei and we compare our findings to widely used parametrizations used in nuclear structure, such as Skyrme energy density functionals (EDFs) or relativistic mean field (RMF)~\cite{Reinhard1995,Bender2003}. 
In particular we discuss the isospin dependence of the spin-orbit interaction. 
Since the present approach can easily be extended to predict the SO interaction for any kind of baryon, we present an application to hyperons in Sec.~\ref{sec:hyp}.
We therefore apply our generic results to the $\Lambda N$ spin-orbit, which is known to be largely quenched, and predict SO potential for the other hyperon systems.
We then conclude this study in Sec.~\ref{sec:conclusions}.

\section{The spin-orbit interaction in atomic and nuclear physics}
\label{sec:so:atom}

The SO interaction exists in many quantum bound systems from atoms to quarkonia, see for instance Ref.~\cite{Ebran2016b}. In atomic physics its origin is well known: it is generated by the coupling of the electron magnetic moment (spin) moving in the electric field of the nucleus, to which shall be added the Thomas precession \cite{Jackson}. An atomic electron -- located at position ${\bf R}$ -- having orbital ${\bf{l}}$ and internal spin ${\bf{s}}$ angular momenta and moving in a central mean-field potential $U(R)$ feels a spin-orbit potential of the form
\begin{eqnarray}
W_{so,e}(R)=\frac{e^2}{m^2_e} \frac{1}{R}\frac{dU}{dR}\,{\bf{l}}\cdot{\bf s}\, -\, \frac{1}{2} \frac{e^2}{m^2_e} \frac{1}{R}\frac{dU}{dR}\,\bf{l}\cdot\bf{s}\,.
\label{eq:so:atom}
\end{eqnarray}
The first term in Eq.~(\ref{eq:so:atom}) comes from the interaction of the electron magnetic moment (represented by the internal spin ${\bf{s}}$) with the mean magnetic field existing in its instantaneous rest frame (IRF): this is a boost effect (generating the gradient) since this mean magnetic field in the IRF originates from the Lorentz transformation of the mean electric field in the rest frame.
However even in the absence of electric and magnetic fields, the rotation of the particle curvilinear orbit involves an additional boost (perpendicular to the motion), which is known as the Thomas precession.
It is a pure relativistic effect that is independent of the structure and that yields the second term in Eq.~(\ref{eq:so:atom}).
The Thomas precession reduces the impact of the first boosted term to the total SO potential.  

The SO interaction in finite nuclei is quantitatively very different.
Not only it is much larger -- see for instance the discussion in Ref.~\cite{Ebran2016b} --
but it has also the opposite sign compared with the atomic physics case. 
While in atomic physics, the coupling of the electromagnetic field to the particle is only of vector type, in nuclear physics, the interaction is spread over more coupling channels. 
In particular, there is also a scalar interaction which contributes to the SO interaction. 
The very large attractive scalar and repulsive vector self-energies, typically $\Sigma_S\approx -400$~MeV and $\Sigma_V\approx+350$~MeV in the interior of finite nuclei, combine together to produce the mean field (sum) and the spin-orbit potential (difference). Consequently the  atomic formula for electrons~(\ref{eq:so:atom}) is transformed for nucleons ($N=p$, $n$) as,
\begin{eqnarray}
W_{so,N}(R)\simeq  \frac{1}{2} \frac{1}{m^2_N} \frac{\Sigma_V - \Sigma_S}{\Sigma_V  + \Sigma_S}\frac{1}{R}\frac{dU}{dR}\,\,\bf{l}\cdot\bf{s}\, .
\label{eq:so:nuc}
\end{eqnarray}
Note that the structure of Eq.~(\ref{eq:so:atom}) can be recovered from Eq.~(\ref{eq:so:nuc}) setting $\Sigma_S=0$.
In Eq.~(\ref{eq:so:nuc}) the nuclear spin-orbit potential is amplified by an order of magnitude since $\Sigma_V - \Sigma_S$ is much larger than $\vert\Sigma_V +\Sigma_S\vert$, and the negative sign is given by the sign of $\Sigma_V + \Sigma_S$.

Moreover in this picture, nucleons in the mean field couple to potentials -- $\Sigma_V$ and $\Sigma_S$ -- which are of the order of one third of their own mass.
One can thus expect that these huge scalar and vector fields probe more than just the global structure factor of nucleons, represented by its mass, but that they are also sensitive to nucleon internal degrees of freedom, such as quarks, gluons, pion cloud, etc... Conversely, it is difficult to imagine that these huge fields have no effect on the internal structure of the nucleon. This is the motivation of the quark-meson coupling (QMC) model proposed by Guichon in Ref.~\cite{Guichon}.
The composite nature of nucleons and the huge fields produces a polarization which softens the density dependence of the energy per particle around saturation density. This mechanism induces a nonlinear sigma-meson coupling or alternatively a density correction to the sigma coupling constant~\cite{Guichon,Chanfray2007,Chanfray2011}.

A question immediately arizes: where do these huge scalar and vector fields come from? How are they generated from QCD and how do they couple to the quarks (and possibly to the pion cloud) inside the nucleon? This will be partially answered in the next section.

\section{Relativistic chiral approach with constraints from nucleon structure}
\label{sec:model}

The link between QCD in its nonperturbative regime and the dynamical interactions among nucleons is not yet completely understood.
Since the spin-orbit interaction between baryons is essentially short-ranged~\cite{Scheerbaum1976}, a number of authors have linked its microscopic origin to the quark degrees of freedom, see for instance the original Ref.~\cite{Pirner1979}. The relation between a quark model and the spin-orbit interaction has also been investigated in the following works, e.g. see Refs~\cite{Guichon, Guichon-so} for nuclei and Refs~\cite{Jennings1990,Cohen1991} for hyper-nuclei. In these models, one usually starts with an effective realization of the low-energy QCD Lagrangian, which can be for instance the Nambu-Jona-Lasinio (NJL) model or nucleon orbital models. 
Recent progresses in Lattice QCD will hopefully help the understanding of the nucleon interaction and the role of the quark substructure.

\subsection{Foundational aspects}

In this section we detail one type of strategy which connects the low-energy realization of QCD and the SO potential.
Here the SO potential, among other things, emerges from a local coupling of vector and scalar fields to the quarks, which are themselves confined by a scalar (string) potential. This can been done in three steps. 

The first step is to perform a gluon averaging of the (euclidean) QCD partition function to generate -- at the so-called Gaussian approximation level -- a chiral invariant four-quark effective Lagrangian. An efficient way is to apply the Field Correlator Method (FCM) elaborated by Y. Simonov an coworkers \cite{Simonov}: a very important outcome is the simultaneous and automatic generation of scalar confinement  and dynamical chiral symmetry breaking. The whole approach depends on two QCD parameters -- the string tension $\sigma$ and the gluon correlation length, or string width, $T_g$ itself related to the gluon condensate -- and yields a long range scalar confining potential $V_C(r)=\sigma r$. What plays the role of a constituent quark mass emerges as  $M\approx \sigma\,T_g$. A possible crude realization but not so bad phenomenologically is the NJL model associated with the following Lagrangian:
\begin{eqnarray}
\mathcal{L} &=& \bar{\psi} \left( i\gamma^\mu\partial_\mu - m \right)\psi + \frac{G_1}{2}\left[ (\bar{\psi} \psi)^2 + (\bar{\psi} i\gamma_5\vec{\tau} \psi)^2\right] \nonumber \\
&&- \frac{G_2}{2}\left[(\bar{\psi} \gamma^\mu\vec{\tau} \psi)^2 + (\bar{\psi} \gamma^\mu\gamma_5\vec{\tau} \psi)^2 + 
(\bar{\psi} \gamma^\mu \psi)^2\right]\, ,
\end{eqnarray}
and complemented by a confining force of a string type~\cite{Chanfray2011}.

In the second step, as explicitly worked out in Ref.~\cite{Chanfray2011}, $q\bar{q}$ fluctuations in the Dirac sea can be integrated out and projected on to mesonic degrees of freedom. This bozonisation procedure generates a scalar field $\sigma$ (with quantum numbers of the "sigma" meson) and vector fields $\omega$, $\rho$ (with quantum numbers of the $\omega$ and $\rho$ mesons) which couple locally to the constituents of the nucleon (the quarks and also possibly the pion cloud which is ignored here). In addition quantum fluctuations generate their kinetic-energy Lagrangian. The model allows us to calculate a quark-scalar and quark-vector coupling constants, $g_{qS}$ and $g_{qV}$, as well as the mass parameters $m_S=m_\sigma$ and $m_V=m_\omega = m_\rho$, which are not the on-shell mesons masses but rather represent the inverse of the corresponding propagators taken at zero momentum. 
According to the FCM approach~\cite{Simonov} and following reference~\cite{Chanfray2011}, the NJL model can be completed by adding a confining force  acting on the NJL constituent quarks whose masses are directly proportional to the in-medium scalar field.

The third step is to evaluate the coupling of these QCD fields  to the nucleon where the quarks move in a scalar confining potential. It thus consists in the emergence of an effective nucleon-meson interaction, as detailed in the next section.

\subsection{From quarks to nucleons}

We now evaluate the coupling of the QCD scalar and vector fields to the nucleons, where constituent quarks move in a (scalar) confining potential. We call generically this type of orbitals model approach "bag model". 
As in Ref. \cite{Guichon-so} let us consider a nucleon at center of mass (CM) position $\bf{R}$ with in-medium effective mass $M_N^*$ and velocity ${\bf{V}}={\bf{P}}/M^*_N$, embedded in the nuclear mesonic fields: $\sigma$, $\omega$ and $\rho$. The quark located at $\bf{R} + \bf{r}$, feels a scalar and a vector potentials:
\begin{eqnarray}
&& U_{qS}\left(\bf{R} + \bf{r}\right)= g_{qS} \,\sigma \left(\bf{R} + \bf{r}\right)\nonumber \\
&&  U_{qV}\left(\bf{R} + \bf{r}\right)= g_{qV} \left(\omega \left(\bf{R} + \bf{r}\right) + \tau_{3q}\, \rho\left(\bf{R} + \bf{r}\right)\right)\, ,
\end{eqnarray}
where $\tau_{3q}$ is the isospin Pauli matrix in the third direction.

The coupling of these mesonic fields to the moving nucleons -- including relativistic effects -- necessitates the knowledge of the three quark wave functions inside the "bag", as discussed in Ref.~\cite{Guichon-so}. 
For that purpose the Instantaneous Rest Frame (IRF) of the nucleon is introduced, with rapidity $\xi$ such that $V=\tanh{\xi}$. 
In the spirit of the Born-Oppenheimer approximation, one can assume that these quarks wave functions are known in the IRF since the quarks have time to adjust their motion so that they are in their lowest energy state (see discussion in  Ref.~\cite{Guichon-so} for details). Performing the boost with rapidity $\xi$ in the IRF of this orbital model, one can define the nucleon mean-field potential as ($N=p$, $n$):
\begin{eqnarray}
 U_N \left(\bf{R}\right)  & = & \int_\mathrm{IRF \,bag} \hspace{-0.5cm} d{\bf{r}}'\,\left\langle N\right| \,\bar{q}\left(\bf{r'}\right)
[U_{qV}\left(\bf{R} + \bf{r'}\right) (\gamma_0 \cosh\xi \nonumber \\
& & +{\bf{\gamma}}\cdot\hat{\bf{V}}\sinh\xi)
+\,U_{qS}\left(\bf{R} + \bf{r'}\right)]\,q\left(\bf{r'}\right)\left|N\right\rangle\, .
\label{MF0}
\end{eqnarray}
We now expand the fields to first and second orders in $\bf{r}'$ according to $U_{qS}\left({\bf{R}} + {\bf{r'}}\right) = U({\bf{R}}) + {{\bf{r}}}'\cdot\vec{\nabla}U\left({\bf{R}}\right)$. Working out the quark-nucleon matrix elements, this generates to leading order the ordinary mean-field potential:
\begin{eqnarray}
U_N ({\bf R}) & \equiv & U_V ({\bf R}) + U_S ({\bf R}) \nonumber \\
&=& g_{\omega} \omega ({\bf R}) \,+ \,g_{\rho}\,\rho({\bf R})\, \left\langle\tau_{3N}\right\rangle  \,+\, g_{\sigma} \,\sigma ({\bf R}).
\end{eqnarray}
where the couplings of the meson fields to the nucleons are defined as $g_\omega= 3\, g_{qV}$, $g_\rho = g_{qV}$, and $g_\sigma = 3\,  g_{qS}\, q_S$, with $q_S =\int d^3r\left(u^2(r)-v^2(r)\right)\,\lesssim 1$ the integrated one-quark scalar density in the nucleon ($u$ and $v$ are the up and down quark radial wave functions in standard notations).
These relations reflect the quark substructure of the nucleon where the factor three refers to the quark number. Let us also mention that the scalar piece of the mean-field potential and the decrease of the nucleon mass in the medium are related to the decrease of the chiral condensate associated with partial chiral symmetry restoration at finite density \cite{Chanfray2007,Chanfray2011}.

Moreover, after performing exactly the inverse boost, this procedure allows us to build a nucleon located at point $\bf{R}$ with energy in the laboratory frame~\cite{Guichon-so},
\begin{equation}
E_0({\bf R}) =\sqrt{M^{*}_N({\bf R})^2 + {\bf P}^{*}({\bf R})^2 }
\end{equation}
with
\begin{equation}
M^*_N ({\bf R}) = M_N + \Sigma_S({\bf R}),\; \hbox{ and }\; {\bf P}^*({\bf R}) = M^*_N ({\bf R}) \hat{{\bf V}}\sinh\xi \, .
\end{equation}

At this level, one comment concerning the coupling to the scalar field is in order. In the detailed approach described in Refs.~\cite{Chanfray2011,Massot2008}, the treatment of the nucleon coupling to the scalar field -- seen as a fluctuation of the chiral field -- is a little more involved since it leads to the concept of the nucleon response to the scalar field as was originally introduced by Guichon \cite{Guichon}. One net effect is the density dependence of the scalar coupling constant corresponding to the progressive reduction of the scalar field, which thus generates the repulsion needed for the saturation mechanism. This mechanism is precisely what was proposed by P. Guichon in his pioneering paper~\cite{Guichon} at the origin of the QMC model. 
Although this scalar field decoupling mechanism is essential for the saturation properties, here we disregard this effect for the SO potential since it is a subleading effect. Another consequence of the scalar nature of the coupling would be to replace in the scalar potential the baryonic density $\rho$ by the scalar density $\rho_S$. This is again a subleading correction which goes beyond the scope of the present study.

For practical applications to nuclear physics our relativistic chiral approach is very similar to the original QMC model~\cite{Guichon2006} but they are at least two important differences at the principle level. First in the QMC model, the nucleon (MIT bag) model only insists on confinement whereas, in our approach, chiral symmetry breaking (and its partial restoration at finite density) is present by construction ; in particular the sigma field has a perfectly-well-defined chiral status; it is chiral invariant and reflects part of the evolution of the quark condensate associated with partial chiral restoration at finite density \cite{Chanfray2007,Chanfray2011}. Second, in the QMC model the local coupling of the three meson fields to the quarks is introduced by hand with six parameters (three coupling constants  and three mass parameters), whereas in our approach, they are generated by the underlying bosonization of the effective QCD Lagrangian.

\subsection{Spin orbit potential}

From Eq.~(\ref{MF0}), the SO potential is defined as the second order in the gradient expansion of the quark position $\bf{r}'$,
\begin{eqnarray}
W_{so,N}({\bf R}) &=& \frac{g_{qV}}{M^*_N({\bf R})} \int_\mathrm{IRF \,bag} \hspace{-0.5cm}d{\bf r}'\, \langle N |\bar{q}({\bf r}') \,\Big[\vec{\bf{\gamma}}\cdot{{\bf P}}\; 
{\bf r}'\cdot \Big(\vec{\nabla}\omega \left({\bf R}\right) \nonumber \\
&&+ \tau_{3q} \vec{\nabla}\rho({\bf R})\Big)\Big]\,q({\bf r}') | N\rangle \, .
\label{eq:SO:pot}
\end{eqnarray}                                                                                
The $\omega$ contribution involves the nucleon matrix element of a one-body quark operator, which can be calculated knowing the up and down quark wave functions:
\begin{eqnarray}
\left\langle g_{qV} \sum_i\,r_i\, \gamma_i\cdot{\bf{P}}\right\rangle &=& -\frac{2}{3}\,g_{qV}\sum_i \left\langle{\bf{\sigma}}_i\times{\bf{P}}\right\rangle \int \!\!d^3r\,r\,u_i(r)\,v_i(r)\,\nonumber \\
&=& -g_\omega\,\frac{2}{9}\, \left\langle{\bf{\sigma}}_N \times{\bf{P}}\right\rangle \int \!\!d^3r\,r\,u(r)\,v(r)\, \nonumber\\
&\equiv& -g_\omega\,\frac{\mu_S}{2\, M_N }\,\left\langle{\bf{\sigma}}_N \times{\bf{P}}\right\rangle , \label{MatrixS}
\end{eqnarray}
and the $\rho$ contribution gives 
\begin{eqnarray}
\left\langle g_{qV} \sum_i\,r_i\, \gamma_i\cdot{\bf{P}}\,\tau_{3i}\right\rangle &=& -\frac{2}{3}\,g_{qV}\,\sum_i  \left\langle{\bf{\sigma}}_i\, \tau_{3i}\times{\bf{P}}\right\rangle  \int \!\!d^3r\,r\,u_i(r)\,v_i(r)\,\nonumber \\
&=& -g_\rho\,\frac{10}{9}\,  \left\langle{\bf{\sigma}}_N \times{\bf{P}}\,\tau_{3N}\right\rangle  \int \!\!d^3r\,r\,u(r)\,v(r)\, \nonumber\\
&\equiv & -g_\rho\,\frac{\mu_V}{2\, M_N }\,\left\langle{\bf{\sigma}}_N \times{\bf{P}}\,\tau_{3N}\right\rangle\, ,\label{MatrixV}
\end{eqnarray}
where in the above equations we used the octet matrix elements,
\begin{eqnarray}
\left\langle N\right|\,\sum_{u,d} {\bf{\sigma}}_i\left|N\right\rangle & = & 
\left\langle N\right|\, {\bf{\sigma}}_N\left|N\right\rangle \, ,\\
\left\langle N\right|\,\sum_{u,d} {\bf{\sigma}}_i\,\tau_{3i}\left|N\right\rangle & = & 
\frac{5}{3}\left\langle N\right|\, {\bf{\sigma}}_N\,\tau_{3N}\left|N\right\rangle\, .\label{SU6}
\end{eqnarray}

The $\omega$ and $\rho$ contributions to the SO potential associated with the boost are directly proportional to the isoscalar and isovector nucleon magnetic moments $\mu_S$ and $\mu_V$, defined in Eqs.~\eqref{MatrixS} and \eqref{MatrixV}, respectively, and calculated in this type of bag model. 
For the SO potential we have implicitly used in the matrix elements the vacuum quark wave functions, ignoring the quark polarization (see above discussion). 
To be consistent the nucleon effective mass is replaced by its free value. 
From Eqs.~(\ref{MatrixS}) and (\ref{MatrixV}), we see that the magnetic moments satisfy the SU(6) bag model ratio $\mu_V/\mu_S=5$ which  is compatible with the values deduced from the experimental neutron and proton magnetic moments: $\mu_V=\mu_p-\mu_n=4.70$ and $\mu_S=\mu_p+\mu_n=0.88$~\cite{PDG}. Moreover it is also possible to explicitly evaluate the radial integral $\int \!\! d^3r\,r\,u(r)\,v(r)$ to get the absolute value of $\mu_S$ and consequently $\mu_p=3\,\mu_S,\, \mu_n=-2\mu_S,\,\mu_V=5\mu_S$. In bag model with a scalar confining interaction the above radial integral can be obtained independently of the precise shape of the confining interaction by using the Dirac equation. The result is
\begin{equation}
\mu_S=\frac{2\,M_N}{9\,\varepsilon_0}\left(1\,+\,\frac{q_S}{2}\right)\equiv\frac{M_N}{10\,\varepsilon_0}\left(g_A\,+\,\frac{5}{3}\right)
\end{equation}
where $\varepsilon_0$ is the eigenenergy of the lowest orbital, $q_S$ is the integrated one-quark scalar density and $g_A$ is the weak axial-vector coupling constant calculated in the model. Considering the experimental value $g_A=1.26$ and taking $M_N\approx 3\,\varepsilon_0$, one automatically obtains the correct order of magnitude, $\mu_S\approx 0.9$. 
However, the MIT bag model predicts $g_A=1.09$ and $\varepsilon_0=2.043/R$, which tends to give too low a magnetic moment for a reasonable value of the bag radius $R\leq 0.8$~fm. In the FCM approach discussed above, chiral symmetry breaking and confinement are simultaneously generated. As already mentioned, this QCD mechanism produces a scalar linear confining interaction, i.e., with Lorentz structure, ${\hat V}_C(r)=\gamma_0\, \sigma r$ where $\sigma$ is the string tension, giving good results for both $g_A$ and the magnetic moments \cite{Weda}.

In the following we introduce the anomalous isoscalar $\kappa_\omega$ and isovector $\kappa_\rho$ magnetic moments through the following definitions: $\mu_S\equiv 1 + \kappa_\omega$ and $\mu_V \equiv 1 + \kappa_\rho$, in units of the nuclear magneton $\mu_N$.

Injecting Eqs.~(\ref{MatrixS}) and (\ref{MatrixV}) into (\ref{eq:SO:pot}), the boost piece of the SO potential takes the following form:
\begin{eqnarray}
W_{so,N}^{boost}({\bf R}) &=& -g_\omega \,\frac{(2 + 2\kappa_\omega)}{4\,M_N M^*_N({\bf R})}\,\vec{\nabla}\omega ({\bf R})\,\cdot\,\left\langle{\bf{\sigma}}_N\times {\bf{P}}\right\rangle \nonumber\\
&&\hspace{-0.5cm}-\,g_\rho\,\frac{(2 + 2\kappa_\rho)}{4\,M_N M^*_N({\bf R})}\, \vec{\nabla}\rho({\bf R})\,\cdot\,
\left\langle{\bf{\sigma}}_N\times {\bf{P}}\,\tau_{3N}\right\rangle\, . \label{eq:Nboost}
\end{eqnarray}

This contribution has to be supplemented by the Thomas precession (TP) piece, also derivable in the above approach \cite{Guichon-so}:
\begin{eqnarray}
W_{so,N}^{TP}({\bf R}) &=& -\,\frac{g_\sigma}{4\,M_N M^*_N({\bf R})}\,\vec{\nabla}\sigma ({\bf R})\,\cdot\,\left\langle{\bf{\sigma}}_N
\times {\bf{P}}\right\rangle \nonumber\\
&& +\,\frac{g_\omega}{4\,M_N M^*_N({\bf R})}\,\vec{\nabla}\omega ({\bf R})\,\cdot\,\left\langle{\bf{\sigma}}_N\times {\bf{P}}\right\rangle \nonumber\\
&&\,+\,\frac{g_\rho}{4\,M_N M^*_N({\bf R})}\, \vec{\nabla}\rho({\bf R})\,\cdot\,
\left\langle{\bf{\sigma}}_N\times {\bf{P}}\,\tau_{3N}\right\rangle\, .
\label{eq:TP}
\end{eqnarray}
Again, for consistency, we also replace the nucleon effective mass coming from the boost by the bare nucleon mass in the following.

Finally, the SO potential is given by the sum of the boost and TP contributions,
\begin{equation}
W_{so,N}({\bf R}) = W_{so,N}^{boost}({\bf R}) + W_{so,N}^{TP}({\bf R}) \, .
\end{equation}

As a side remark, let us mention that the above results can also be derived from a relativistic theory such as that utilized in Ref.~\cite{Massot2008} where the nucleon-vector meson coupling Lagrangian written with standard notation reads:
\begin{eqnarray}
{\cal L}_\omega &=& - g_\omega\,\omega_\mu\,\bar\Psi\gamma^\mu\Psi\,-\,g_\omega\frac{\kappa_\omega}{2\,M_N}\,\partial_\nu \omega_{\mu}\,\Psi\bar\sigma^{\mu\nu}\Psi \, ,\nonumber\\	
{\cal L}_\rho &=&- g_\rho\,\rho_{a\mu}\,\bar\Psi\gamma^\mu \tau_a\Psi
\,-\,g_\rho\frac{\kappa_\rho}{2\,M_N}\,\partial_\nu \rho_{a\mu}\,\Psi\bar\sigma^{\mu\nu}_{}\tau_a\Psi\, . \nonumber \\
\end{eqnarray}
The origin of the tensor ($\rho$ and $\omega$) couplings in such a Lagrangian is not resolved but instead given as an input. In our approach instead, these couplings are derived from the quark substructure of the baryons.

\section{Spin-orbit potential in nuclei}
\label{sec:nuc}

The equations of motion for the meson fields can be used to express the SO potential in terms of the nucleon densities. 
Starting from Eqs.~(\ref{eq:Nboost}) and (\ref{eq:TP}) and assuming large vector-meson masses (i.e., neglecting Darwin terms), e.g. $\omega \left({\bf{R}}\right)=\frac{g^2_\omega}{m^2_\omega}n_0(r)$, we obtain, after elementary manipulations (namely, ${\bf R}\cdot\,\left\langle{\bf{\sigma}}_N\times {\bf{P}}\right\rangle =-2\left(\bf{l}\cdot\bf{s}\right)$), an expression for spherical nuclei involving the radial derivative of the total nucleon density $n_0(r) \equiv n_p(r) +n_n(r)$ and the isovector density $n_1(r) \equiv n_p(r) - n_n(r)$ (note the convention for $n_1$ which is opposite to the usual nuclear one) as,
\begin{eqnarray}
W_{so,N}^{boost}(R)&= &\frac{1}{2\,R\,M^2_N}\Big[\frac{g^2_\omega}{m^2_\omega}(2 + 2\kappa_\omega)\frac{dn_0}{dR}\nonumber\\
&&\pm\frac{g^2_\rho}{m^2_\rho} (2 + 2\kappa_\rho)\frac{dn_1}{dR}\Big] \left(\bf{l}\cdot\bf{s}\right)_{\tau}\, .
\end{eqnarray}
and
\begin{equation}
W_{so,N}^{TP}(R)=\frac{1}{2\,R\,M^2_N}\Big[\frac{g^2_\sigma}{m^2_\sigma}\frac{dn_0}{dR}-\frac{g^2_\omega}{m^2_\omega}\frac{dn_0}{dR}
\mp\frac{g^2_\rho}{m^2_\rho} \frac{dn_1}{dR}\Big] \left(\bf{l}\cdot\bf{s}\right)_{\tau} \, .
\end{equation}
where the  $\pm$ and $\mp$  signs in the previous equations refer respectively to the proton and neutron cases.

In practice, due to the small value of $\kappa_\omega\sim -0.13$, the $\omega$-tensor coupling can be safely neglected.
The case of the $\rho$ meson is less simple.
The pure Vector Dominance Model (VDM) picture \cite{Bhaduri}, i.e., the strict proportionality between the electromagnetic current and the vector-meson fields, implies the identification of $\kappa_{\rho}$ with the anomalous part of the isovector magnetic moment of the nucleon, {\it i.e.}, $\kappa_{\rho}=3.7$, hereafter called the medium coupling for the $\rho$ meson. 
For instance the effective Lagrangian PKA1~\cite{LongPKA1} assumes a value of about 3.2, which is comparable to the one suggested by the VDM picture.
However pion-nucleon scattering data~\cite{HP75} suggest a larger value $\kappa_{\rho}=6.6$, hereafter called the strong $\rho$ coupling.
Many approaches in finite nuclei while including the $\rho$-vector coupling neglect the $\rho$-tensor coupling; see, for instance, Ref.~\cite{LongPKO,Ebran2016}.
In the following, we also explore this case, defined as the weak $\rho$ coupling. 
Finally, the decomposition of the SO potential for the various meson channels are shown in Table~\ref{tab:LS:NN}.

\begin{table}[t]
\centering
\setlength{\tabcolsep}{5pt}
\renewcommand{\arraystretch}{1.5}
\begin{tabular}{cccccccccccc}
\hline
Meson & Boost & Thomas & Total & Associated\\
 & & precession & & gradient \\
\hline
$w_{N}^\sigma$ & 0 & $\frac{g_\sigma^2}{m_\sigma^2}$ & $\frac{g_\sigma^2}{m_\sigma^2}$ & $\nabla n_0$  \\
$w_{N}^\omega$ & $2(1+\kappa_\omega) \frac{g_\omega^2}{m_\omega^2}$ & $-\frac{g_\omega^2}{m_\omega^2}$ & $(1+2\kappa_\omega) \frac{g_\omega^2}{m_\omega^2}$ & $\nabla n_0$\\
$w_{N}^\rho$ & $2(1+\kappa_\rho) \frac{g_\rho^2}{m_\rho^2}$ & $-\frac{g_\rho^2}{m_\rho^2}$ & $(1+2\kappa_\rho) \frac{g_\rho^2}{m_\rho^2}$ & $\nabla n_1$\\
\hline
\end{tabular}
\caption{Meson decomposition of the nucleon direct (Hartree) spin-orbit potential multiplying the term 
$\left(\bf{l}\cdot\bf{s}\right)_{\tau}/(2 R M_N^2)$. The densities are $n_0\equiv n_n+n_p$ and $n_1\equiv n_p-n_n$.}
\label{tab:LS:NN}
\end{table}

For the nucleonic sector, the SO potential is usually expressed as,
\begin{equation}
W_{so,N}({\bf R}) = \frac{1}{R}\left( W_1 \nabla n_\tau + W_2 \nabla n_{-\tau}\right) ({\bf{l}}\cdot{\bf{s}})_\tau \, ,
\label{eq:wso12}
\end{equation}
hence directly exhibiting its isospin dependence. 
For the Skyrme interaction, the ratio $(W_1/W_2)^\mathrm{Skyrme}=2$ and for Walecka-type RMF models (without the $\rho$) we have $(W_1/W_2)^\mathrm{RMF,\,no\,\rho}=1$~\cite{Reinhard1995,Bender2003}.

One could express the coefficients $W_1^H$ and $W_2^H$ for the direct (Hartree) contribution in terms of the quantities $w_{N}^i$ ($i=\sigma$, $\omega$, $\rho$) defined in Table~\ref{tab:LS:NN},
\begin{eqnarray}
W_1^H &\equiv& \frac{1}{2M_N^2} \left[ w_{N}^\sigma+ w_{N}^\omega+w_{N}^\rho \right] \, ,\label{eq:H:1}\\
W_2^H &\equiv& \frac{1}{2M_N^2} \left[ w_{N}^\sigma+ w_{N}^\omega-w_{N}^\rho \right] \, .\label{eq:H:2}
\end{eqnarray}

For the orientation of the following discussion let us reasonably consider that the $\sigma$ and the $\omega$ contributions are similar, as suggested from most phenomenological studies~\cite{Walecka1997,Bender2003}. 
For instance if we choose the omega coupling adjusted from standard VDM phenomenology ($g_{qV}=2.65$, $m_\omega=780$~MeV)~\cite{Bhaduri}, one obtains ${g_\omega}/{m_\omega}= (3\times 2.65\times 200/780)\simeq 2$~fm.
Note that the effective Lagrangian PKA1~\cite{LongPKA1} directly calibrated from nuclear properties suggests ${g_\omega}/{m_\omega}=2.7$~fm, which is not far from our current estimate, considering that the PKA1 Lagrangian is obtained by ignoring the quark substructure.
To obtain the binding of nuclear matter, ${g_\sigma}/{m_\sigma}$ shall be slightly larger, leading to $w_{N}^\sigma\sim 1.1\, w_{N}^\omega$ at maximum. 
The $\rho$ coupling constant is one third of the $\omega$ coupling constant.
So in the absence of $\rho$-tensor coupling (weak $\rho$), we also expect  that the $\rho$ contribution will be $w_{N}^{\mathrm{weak\, \rho}}\sim 0.1\, w_{N}^\omega$. 
However in the medium- and strong-$\rho$ cases, $w_{N}^{\mathrm{med\,\rho}}$ and $w_{N}^{\mathrm{strong\,\rho}}$ will be around ten times as large; see Table~\ref{tab:NNcases} for typical values.

\begin{table}[t]
\centering
\setlength{\tabcolsep}{7pt}
\renewcommand{\arraystretch}{1.5}
\begin{tabular}{ccccc}
\hline
$\rho$ contribution & no $\rho$ & weak $\rho$ & medium $\rho$ & strong $\rho$ \\
\hline
$w_{N}^\rho$ (fm$^{2}$) & 0 & $\simeq 0.4$ & $\simeq 3.2$ & $\simeq 5.6$ \\
\hline
$W_1^H/W_2^H$ & 1 & $\simeq 1.1$  & $\lesssim 3$ & $\gtrsim 3$\\
$W_1^{HF}/W_2^{HF}$ & 1.5 & $\simeq 1.7$ & $\lesssim 2.25$ & $\gtrsim 2.25$ \\
\hline
$\frac 1 2 (W_1 + W_2)^{H}$ & $\simeq 0.18$ & $\simeq 0.18$ & $\simeq 0.18$ & $\simeq 0.18$  \\
(fm$^{4}$) & \\
$\frac 1 2 (W_1 + W_2)^{HF}$ & $\simeq 0.23$ & $\simeq 0.25$ & $\simeq 0.35$ & $\simeq 0.42$ \\
(fm$^{4}$) & \\
\hline
\end{tabular}
\caption{Summary of the results showing the various scenarios for the $\rho$.
First row, $\rho$ coupling $w_{N}^\rho$ for the weak, medium and strong scenarios.
For the other mesons, we have $w_{N}^\omega\simeq 4$~fm$^{2}$ and $w_{N}^\sigma\simeq 4.4$~fm$^{2}$.
Then we show the predictions for the ratio $W_1/W_2$, see Eqs.~(\ref{eq:H:1}), (\ref{eq:H:2}), (\ref{eq:NNWHF}), and for the half sum $(W_1+W_2)/2$; see Eq.~(\ref{eq:NNW1W2sum}), for Hartree (RMF) and Hartree-Fock (RHF) cases.}
\label{tab:NNcases}
\end{table}

Hence for a typical RMF approach without the $\rho$ field, Eqs.~(\ref{eq:H:1}) and (\ref{eq:H:2}) predict $W_1^H/W_2^H=1$, as expected from the simplest "$\sigma$-$\omega$" Walecka model; see, for instance Ref.~\cite{Reinhard1995}.
It is interesting to note that, for the medium (strong) $\rho$ coupling, one has  $W_1^H/W_2^H\lesssim 3$ ($W_1^H/W_2^H\gtrsim 3$) significantly larger than $W_1^H/W_2^H\simeq 1.1$ for the weak $\rho$ coupling.
In Ref.~\cite{Ebran2016} this ratio calculated at the Hartree level (RMF) remains very close to 1.1 (see Fig.~1 of this paper) for various nuclei ($^{16}$O, $^{34}$Si, $^{208}$Pb), which is consistent with the weak-$\rho$ hypothesis.

We see the considerable effect of the $\rho$-tensor coupling to the isovector density dependence of the SO potential, which in our approach is interpreted as a purely quark substructure effect. 
The question then arises of its survival when exchange terms are included. 
From the approach of Ref.~\cite{Massot2008}, it is possible to construct  an energy density functional~\cite{Chanfray2019} as in the QMC model~\cite{Guichon2006,Stone2016}, from which one can deduce the exchange (Fock) contribution to the spin-orbit potential. 
The results for the direct (Hartree) are given in Eqs.~(\ref{eq:H:1}) and (\ref{eq:H:2}), and we give hereafter the exchange (Fock) and total (Hartree-Fock) contribution to $W_1$ and $W_2$ as,
\begin{eqnarray}
W_1^F &=& \frac{1}{2M_N^2} \frac 1 2 \left(w_{N}^\sigma + w_{N}^\omega+w_{N}^\rho\right)\, , \nonumber \\
W_2^F &=& \frac{1}{2M_N^2} w_{N}^\rho\, , \nonumber \\
W_1^{HF} &=& \frac{1}{2M_N^2} \frac 3 2 \left(w_{N}^\sigma+ w_{N}^\omega+w_{N}^\rho\right) \, , \nonumber \\
W_2^{HF} &=& \frac{1}{2M_N^2} \left( w_{N}^\sigma+ w_{N}^\omega \right) \, .
\label{eq:NNWHF}
\end{eqnarray}
Completely ignoring the contribution of the $\rho$ meson, one gets $\left[W_1^{HF}/W_2^{HF}\right]^\mathrm{no\,\rho} = 1.5$, which can be seen as the basic Walecka model result with exchange correction included. 
Introducing the $\rho$ contribution in the weak scenario, i.e., ignoring the tensor coupling, one obtains $\left[W_1^{HF}/W_2^{HF}\right]^\mathrm{weak\,\rho}\sim 1.7$, not far from the ratio $1.75-1.8$ visible in Fig.~2 of Ref.~\cite{Ebran2016} for the interior of $^{16}$O, $^{34}$Si, $^{208}$Pb, in a RHF calculation  which  also incorporates density dependent couplings. 
For the medium $\rho$ coupling, the ratio becomes $\left[W_1^{HF}/W_2^{HF}\right]^\mathrm{med\,\rho}\lesssim 2.25$, and for the strong $\rho$ coupling, it becomes $\left[W_1^{HF}/W_2^{HF}\right]^\mathrm{strong\,\rho}\gtrsim 2.25$. 
One can also compare with the conventional Skyrme EDF parametrization for which $W_1/W_2=2$, see for instance Refs.~\cite{Reinhard1995,Bender2003}, which turns out to be close to our estimate for medium and strong $\rho$ scenarios. 
These results are summarized in Table~\ref{tab:NNcases}.

One can observe from the results given in Table~\ref{tab:NNcases} that the ratio $W_1/W_2$ is clearly influenced by the contribution of the $\rho$ meson, as well as by the Fock term in the mean field. 
For all cases, the Fock term contributes to shift the ratio $W_1/W_2$ towards the phenomenological Skyrme value ($\simeq 2$).
A systematical analysis based on experimental data, e.g., see the comparisons in Refs.~\cite{Bender2003,Li2014}, can provide a clear insight on the strength of the $\rho$ meson coupling.

It is also interesting to look at the influence of the $\rho$ meson on the absolute value of the SO potential. 
For this purpose one can look at its isoscalar component, i.e., the SO potential felt by one nucleon in a $N=Z$ nucleus.
In the SLy5 Skyrme EDF approach this potential is parametrized with the $W_0\simeq 120$~MeV~fm$^{5}\simeq 0.6$~fm$^{4}$ parameter~\cite{Chabanat} according to:
\begin{eqnarray}
\left[W_{so}\right]^\mathrm{Skyrme}_{N=Z}(R) = \frac 3 4 \frac{W_0}{R}\, \frac{dn_0}{dR} \, {\bf{l}}\cdot{\bf{s}} .
\end{eqnarray}
In our microscopic approach, the same quantity is given by: 
\begin{eqnarray}
\left[W_{so}\right]^\mathrm{Micro}_{N=Z}(R) = \frac{ W_1 + W_2}{2 R}\, \frac{dn_0}{dR} \, {\bf{l}}\cdot{\bf{s}} .
\end{eqnarray}
We see that we have to compare $3 W_0/4\simeq 0.45$~fm$^4$ in the SLy5 Skyrme EDF with $(W_1+W_2)/2$ from the microscopic approach, which is expressed as,
\begin{eqnarray}
&&\left[\frac{W_1 + W_2}{2}\right]^{H}_\mathrm{no\, \rho} = \frac{1}{2M_N^2} \left( w_{N}^\sigma + w_{N}^\omega \right) \nonumber\, ,\\
&&\left[\frac{W_1 + W_2}{2}\right]^{H}_\mathrm{with\, \rho} = \frac{1}{2M_N^2} \left( w_{N}^\sigma + w_{N}^\omega \right) \nonumber\, ,\\
&&\left[\frac{W_1 + W_2}{2}\right]^{HF}_\mathrm{no\, \rho} = \frac{5}{8M_N^2} \left(w_{N}^\sigma + w_{N}^\omega\right) \nonumber\, ,\\
&&\left[\frac{W_1 + W_2}{2}\right]^{HF}_\mathrm{with\, \rho} = \frac{5}{8M_N^2} \left(w_{N}^\sigma + w_{N}^\omega\right) + \frac{3}{8M_N^2} w_{N}^\rho \, .
\label{eq:NNW1W2sum}
\end{eqnarray}

The two last rows of Table~\ref{tab:NNcases} provide estimates for the SO potential under various scenarios for the $\rho$ coupling.
We see that the contribution of the $\rho$ meson, including its tensor piece, is of utmost importance to reproduce the Skyrme phenomenology otherwise the SO potential would be strongly underestimated by almost a factor of two. 
Also note that the quantitative agreement of the microscopic approach with the Skyrme interaction has been discussed within the QMC model in Ref.~\cite{Stone2016}.
Moreover the strength of $\rho$-tensor coupling, which is still under discussion, can possibly be determined from the isoscalar and isovector density dependence of the SO interaction extracted from finite nuclei data.

Let us make a further comment: in the present case we have neglected the influence of the nucleon effective mass, namely its lowering with respect to the bare nucleon mass which is position dependent and varies from about 0.6-0.7$M_N$ in the bulk up to about $M_N$ at the surface. At leading order, the SO potential is increased by about 10\%-30\% up, see Eqs.~\eqref{eq:Nboost} and \eqref{eq:TP}, depending on the coordinate position ${\bf R}$; to be consistent one should nevertheless take into account the decrease of the in-medium scalar coupling constant $g_\sigma$, which reduces the SO potential but to a lower extent. As a result, the SO potential will increase by about 10\%-20\% while the ratio $W_1/W_2$ will be almost unchanged. The Skyrme phenomenology will thus be recovered more consistently for the isoscalar and isovector density dependence by considering the medium-to-strong-coupling cases.
These results certainly deserve a more detailed calculation but a firm conclusion is nevertheless that a realistic relativistic calculation (RHF) certainly requires the inclusion of the $\rho$-tensor coupling, which can ultimately be linked to the quark substructure of the nucleon.
The symmetry (SLS) and the antisymmetric (ALS) spin-orbit terms to the energy splitting are discussed in Refs.~\cite{Hiyama2000,Vesely2016}.

\section{Spin-orbit potential in hyper-nuclei}
\label{sec:hyp}

Let us now come to the question of the SO potential in hyper-nuclei. Recent precision measurements of $E1$ transitions from $p-$ to $s-$shell orbitals of a $\Lambda$ hyperon in $^{13}_\Lambda$C give a $p_{3/2}$-$p_{1/2}$ SO splitting of only (152$\pm$65)~keV~\cite{Ajimura2001} to be compared with about 6~MeV in ordinary $p-$shell nuclei (different by a factor $\approx 50$). The $\Lambda$ SO potential therefore appears to be weaker by at least an order of magnitude than the nucleonic SO potential.
This effect was originally suggested from phenomenological analyses indicating a strong suppression of the $\Lambda$ spin-orbit potential~\cite{Bouyssy1976,Bouyssy1980}.

Since the seminal work by Brockmann and Weise~\cite{Brockmann1977} where the reduction of the SO potential in $\Lambda$ hypernuclei was obtained in a relativistic Hartree approach, this effect has been investigated within several models: From one-boson exchange $N\Lambda$ potentials~\cite{Dover1984,Rijken1999,Hiyama2000,Millener2001}, which tend to overestimate the $N\Lambda$ spin-orbit potential; from SU(3) generalization of standard nuclear RMF models~\cite{Brockmann1977,Bouyssy1982,Papazoglou1999,Muller2000}; from the naive SU(6) quark model with flavor symmetry breaking, which naturally explains the small spin-orbit coupling of the $\Lambda$ hyperon; from a quark model picture combined to Dirac phenomenology~\cite{Jennings1990, Chiapparini1991}; or from combining the quark model with scalar- and vector-meson exchange (QMC, quark-meson coupling model)~\cite{Tsushima1998}, where Pauli blocking in the $\Lambda N$-$\Sigma N$ coupled channels is incorporated phenomenologically.
We finally mention the flavor-SU(3) in-medium chiral effective-field theory approaches, where strangeness is being included. An almost complete cancellation is found between short-range contributions and long-range terms~\cite{Kaiser2003,Kaiser2005}, even including the three-body spin-orbit interaction of Fujita-Miyazawa type~\cite{Kaiser2008}. This scenario has been tested over a large set of hyper-nuclei~\cite{Finelli2007,Finelli2009}.
This list is only partial and many other approaches have been developed. 

\begin{table}[t]
\centering
\setlength{\tabcolsep}{8pt}
\renewcommand{\arraystretch}{1.5}
\begin{tabular}{cccccc}
\hline
Baryon & composition & $S_B$ & $T_B$ & $L_B$ & $I_B$  \\
\hline
$p$   & uud & $1$  & $5/3$ & $1$  &  $1$  \\
$n$   & udd & $1$  & $-5/3$ & $1$  &  $-1$  \\
$\Lambda$ & uds & $0$ & $0$ & $2/3$ & $0$ \\
$\Sigma^+$ & uus & $4/3$ & $4/3$ & $2/3$ & $2$ \\
$\Sigma^0$ & uds & $4/3$ & $0$ & $2/3$ & $0$ \\
$\Sigma^-$ & dds & $4/3$ & $-4/3$ & $2/3$ & $-2$ \\
$\Xi^0$ & uss & $-1/3$ & $-1/3$ & $1/3$ & $1$ \\
$\Xi^-$ & dss & $-1/3$ & $1/3$ & $1/3$ & $-1$ \\
$\Omega^-$ & sss & $0$ & $0$ & $0$ & $0$ \\
\hline
\end{tabular}
\caption{Matrix elements for nucleons and hyperons.}
\label{tab:matrixelements}
\end{table}

Our chiral relativistic approach can be extended to the full octet including hyperons. 
In case of a single hyperon hypernucleus, the mesonic mean field originating from the  ensemble of baryons with a large majority of nucleons is not modified. When considering the spin-orbit felt by an hyperon, the summation on the quarks appearing in Eqs.~(\ref{MatrixS}) and (\ref{MatrixV}) has to be limited to the $u$ and $d$ quarks since the strange quark does not couple to the $\sigma$, $\omega$ and $\rho$ fields.
We neglect here interaction of hyperons mediated by strange mesons. 
Assuming SU(3) flavor symmetry, one can derive a general expression for the SO potential experienced by any baryon $B=N$ or $Y$ where $Y=\Lambda$, $\Xi$, $\Sigma$ or $\Omega$~\cite{Tsushima1998}:
\begin{equation}
W_{so,B}(R) = W_{so,B}^{Boost}(R)+W_{so,B}^{TP}(R) \, ,
\label{eq:wsob}
\end{equation}
where
\begin{eqnarray}
W_{so,B}^{Boost}(R) &=& \frac{1}{2 R M_N^2}\left[\frac{g_\omega^2}{m_\omega^2}\left(2+2\kappa_\omega\right)\,S_B\,\frac{dn_0}{dR}\right. \nonumber \\
&&\hspace{1cm} +\,\left. \frac{g^2_\rho}{m^2_\rho}\left(2+2\kappa_\rho\right)\,\frac{3}{5}\,T_B \frac{dn_1}{dR}\right]\,\left(\bf{l}\cdot\bf{s}\right)_B 
\label{eq:BSO:boost}\\
W_{so,B}^{TP}(R) &=& \frac{1}{2 R M_N^2}\left[\left(\frac{g_\sigma^2}{m_\sigma^2}\,-\,\frac{g_\omega^2}{m_\omega^2}\right)\,L_B\,\frac{dn_0}{dR}\,-\,\frac{g^2_\rho}{m^2_\rho}\,I_B \frac{dn_1}{dR}\right]\,\left(\bf{l}\cdot\bf{s}\right)_B\nonumber \\
\label{eq:BSO:TP}
\end{eqnarray}
with
\begin{eqnarray}
S_B &=&\frac{\left\langle \,\sum_{i=u,d} \,\sigma_{3i}\, \right\rangle_B}{\left\langle \,\sigma_{3B}\, \right\rangle_B}\, , \,\,
T_B=\frac{\left\langle \,\sum_{i=u,d} \,\sigma_{3i}\,\tau_{3i}\, \right\rangle_B}{\left\langle \,\sigma_{3B}\, \right\rangle_B}\, , \\
L_B &=& \left\langle \frac{1}{3}\,\sum_{i=u,d} \,1_i\, \right\rangle_B \, , \,\,
I_B = \left\langle \,\sum_{i=u,d} \,\tau_{3i}\, \right\rangle_B \, ,
\end{eqnarray}
where only the Hartree term is considered since we treat the single hyperon case.
The relevant SU(6) matrix elements $S_B$, $T_B$, $L_B$ and $I_B$ are given in Table~\ref{tab:matrixelements}.

A first general remark is that the Thomas precession -- which was already small for the nucleon SO potential due to the compensation between the scalar and the vector terms, and the small contribution of the $\rho$ term -- is also small for the hyperon potential for the same reason; see Eq.~(\ref{eq:BSO:TP}).

Let us give explicitly the SO potential for the neutral hyperons, namely $\Lambda$, $\Sigma^0$ and $\Xi^0$. To simplify the writing we omit the prefactor $1/2 R M_N^2$ and we define $G_\sigma=g_\sigma^2/m_\sigma^2$, $G_\omega=g_\omega^2/m_\omega^2$, $G_\rho=g_\rho^2/m_\rho^2$, also keeping in mind that $G_\sigma\approx G_\omega\approx (1+2\kappa_\rho) G_\rho\approx 10\, G_\rho$ in the case of medium and strong $\rho$ couplings.
\begin{equation}
W_{so,\Lambda}=\frac{2}{3}\left(G_\sigma-G_\omega\right)\frac{dn_0}{dR}\left(\bf{l}\cdot\bf{s}\right)_\Lambda\label{SOLambda}
\end{equation}
\begin{equation}
W_{so,\Sigma^0}=\left[\frac{2}{3}\left(G_\sigma-G_\omega\right)+\frac{4}{3}G_\omega\left(2+2\kappa_\omega\right)\right]\frac{dn_0}{dR}\left(\bf{l}\cdot\bf{s}\right)_{\Sigma^0}\label{SOSigma}
\end{equation}

\begin{eqnarray}
W_{so,\Xi^0} &=& \left\{\left[\frac{1}{3}\left(G_\sigma-G_\omega\right)-\frac{1}{3}G_\omega\left(2+2\kappa_\omega\right)\right]\,\frac{dn_0}{dR} \right.\nonumber \\  
&& -\,\left.\left[G_\rho +\frac{1}{5}G_\rho\left(2+2\kappa_\rho\right)\right] \frac{dn_1}{dR}\right\}\,\left(\bf{l}\cdot\bf{s}\right)_{\Xi^0} 
\label{eq:SOXi}
\end{eqnarray}
to be compared with the SO potential for neutrons, including here the Fock contribution,
\begin{eqnarray}
W_{so,n}&=&\left[\frac{W_1 + W_2}{2} \,\frac{dn_0}{dR}-\frac{W_1 - W_2}{2}\, \frac{dn_1}{R}\right]\,\left(\bf{l}\cdot\bf{s}\right)_n 
\nonumber\\
 &=& \left\{\left[\frac{5}{4}\left(G_\sigma+G_\omega\left(1+2\kappa_\omega\right)\right)+\frac{3}{4}G_\rho\left(1+2\kappa_\rho\right)\right]\,\frac{dn_0}{dR}\right.\nonumber \\  
&-& \left.\left[G_\rho\left(1+2\kappa_\rho \right)+\frac{1}{4}\left(G_\sigma+G_\omega\left(1+2\kappa_\omega\right)\right)\right] \frac{dn_1}{dR}\right\}\,\left(\bf{l}\cdot\bf{s}\right)_{n} \nonumber \\
\label{eq:SOneutron}
\end{eqnarray}

In the particular case of the $\Lambda$ hyperon, Eq.~(\ref{SOSigma}), Thomas precession is however the only term which survives, leading to a strong reduction of the SO potential (by a factor of about 50, as in the experimental data) with respect to the neutron case.
Similar conclusions have also been obtained in various analyses, see Refs.~\cite{Bouyssy1976,Bouyssy1980,Jennings1990,Tsushima1998}. 

Concerning the $\Sigma$ case, Brockmann~\cite{Brockmann1981} has predicted a small spin-orbit splitting, in contrast with quark-model predictions suggesting a strong spin-orbit splitting~\cite{Pirner1979}. In our case we predict for symmetric nuclei an increase of the $\Sigma$ SO potential by about thirty percent with respect to the neutron case taken at the Hartree level, for N=Z, i.e., $^{41}_{\Sigma^0}$Ca as a typical example.  However, if the Fock term is taken into account for the neutron case the Hartree term is increased by a factor of $5/4$ to which a $\rho$ contribution has to be added. Hence, for the weak-$\rho$ coupling, the $\Sigma$ SO potential is expected to be very close to the neutron case whereas, for the medium- or strong-$\rho$ coupling the $\Sigma$ SO potential is expected to be twenty percent smaller than the neutron SO potential.  The QMC model (see Fig.~3 of Ref.~\cite{Tsushima1998}) predicts a slight decrease in the case of $^{41}_{\Sigma^0}$Ca of this order of magnitude. 
We also observe that the $\Sigma$ SO potential is dominated by the $\omega$ meson, which induces a great stability in our results, almost independent of the $\rho$ scenario.

For the cascade case we predict, again for symmetric nuclei, a significant reduction by a factor of one fifth with respect to the neutron case with in addition a change of sign (also observed in Ref.~\cite{Tsushima1998}) . One peculiarity of the $\Xi$ SO potential is that the contribution of the $\omega$ meson is quenched. In asymmetric nuclei such as $^{209}_{Y}$Pb, the reduction of the hyperon spin-orbit potential compared with the neutron spin-orbit potential is even accentuated. 

Finally, for the $\Omega$ hyperon there is no SO potential since its is composed only of strange quarks.

So, in conclusion, we predict very different SO potentials for hyperons based on different meson coupling mechanisms: the cancellation of the boost contribution strongly quenches the $\Lambda$ SO potential, the dominance of the $\omega$ coupling for the $\Sigma$ SO potential induces a very stable prediction, which is 20\% smaller than for nucleons, and finally, the quenching of the $\omega$ contribution for the $\Xi$ SO potential makes smaller by a factor of order five depending on the $\rho$ scenario compared with nucleons with in addition a change of sign.

\section{Conclusions}
\label{sec:conclusions}

In this paper, the predictions of a chiral relativistic approach for the SO potential in nuclei and hyper-nuclei are analyzed. The basic inputs are introduced  at the quark substructure level  in such a way that the 
$\omega$ and $\rho$ coupling constants are compatible with  the standard VDM phenomenology and that the $\sigma$ coupling allows a plausible saturation mechanism. The strength of the anomalous magnetic moment generated from the $\rho$-tensor coupling is also predicted from the VDM picture, and we explore some departure from it. Specifically, we study three distinct scenarios: the weak-coupling case (no $\rho$ tensor), the medium coupling case (suggested from quark substructure and VDM), and the strong-coupling case (deduced from  pion-nucleon scattering data). In finite nuclei, the important role of the $\rho$ meson is underlined and we compare our results to the usual approximations, where either the $\rho$ meson is neglected or the Fock term is not calculated (as in the RMF approach).
We show that the systematics of SO splitting in finite nuclei could be used to better determine the strength of the $\rho$ meson coupling.
An important result is that the Skyrme phenomenology can be recovered only in the case of the medium to strong $\rho$ coupling.

The present model is based on the quark substructure of the nucleon, sensitive both to the confinement mechanism and to SU(3) symmetry for the values or  relations between the $\sigma$, $\omega$ and $\rho$ mesons coupling constants. The strong sensitivity of the results on the $\rho$ meson strength -- and in particular on its tensor piece affecting the nucleon anomalous magnetic moment -- suggests that the confrontation of the present phenomenological analysis for systematics in finite nuclei could shed light on the quark substructure of nucleons.

The same chiral relativistic model is applied to hypernuclei where is shows that the SO potential in these cases can be very different. It is quenched for $\Lambda$, decreased by 20\% for the $\Sigma$, and reduced by one fifth for the $\Xi$ case. For each case, the mechanism is different and analyzed in the present approach.

Extending the present model to the description of finite nuclei, it will be interesting to analyze the isotope shifts in the Pb region since it is expected to be closely related to the spin-orbit interaction as well~\cite{Reinhard1995}.
In the future, we should include other contributions, such as the $\pi$ and $\delta$ mesons missing in the present approach.

\end{document}